\documentclass[12 pt]{article}
\usepackage{times}
\usepackage{graphicx}
\usepackage{bm}
\usepackage[pagebackref=false,colorlinks,linkcolor=blue,citecolor=blue]{hyperref}
\usepackage[margin=1.1 in]{geometry}
\usepackage{booktabs} 
\usepackage{array} 
\usepackage{paralist} 
\usepackage{verbatim} 
\usepackage{subfig} 
\usepackage{fancyhdr} 
\usepackage{sectsty}
\usepackage[nottoc,notlof,notlot]{tocbibind} 
\usepackage[titles,subfigure]{tocloft} 
\usepackage{cite}
\usepackage{amsmath}
\usepackage[usenames, dvipsnames]{color}
\usepackage{float}
\usepackage[affil-it,]{authblk}
\begin{document}

\title{ \emph{Q-deformed Morse and Oscillator potential }}

\author[1]{\textit{H Hassanabadi}%
	\thanks{Electronic address:h.hasanabadi@shahroodut.ac.ir (Corresponding author)}}
\affil[1]{{\small Department of Physics , Shahrood University of Technology, Shahrood, Iran \\ P. O. Box: 3619995161-316.}}

\author[2]{\textit{W S Chung}}
\affil[2]{{\small Department of Physics and Research Institute of Natural Science,
		College of Natural Science,
		Gyeongsang National University, Jinju 660-701, Korea. }}
\author[3]{\textit{S Zare}}
\affil[3]{{\small Department of Basic Sciences, Islamic Azad University, North Tehran Branch, Tehran, Iran. }}
\author[4]{\textit{S B Bhardwaj}}
\affil[4]{{\small Department of Physics, Kurukshetra University, Kurukshetra 136 119, India. }}

\maketitle

\begin{abstract}
In the present work, we studied the q-deformed Morse and harmonic oscillator systems with appropriate canonical commutation algebra. The analytic solutions for eigenfunctions and energy eigenvalues are worked out using time-independent Schr\"odinger equation and It is also noted that these wave functions are sensitive to variation in the parameters involved.
\end{abstract}

\begin{small}
\end{small}\\

\begin{small}
\end{small}
\\
\newpage
\section{Introduction}

Quantum groups and q-deformed algebras have been the subject of intense study and investigation in the last decade. In the past few years, a q-deformed harmonic oscillator was introduced \cite{1,2,3} and then inspiring of such deformation, quantum groups and q-deformations have found applications in various branches of physics and chemistry, specially, they have been utilized to express electronic conductance in disordered metals and doped semi conductors \cite{4}, to analyze the phonon spectrum in $^4$He \cite{5}, to specify the oscillatory-rotational spectra of diatomic \cite{6} and multi-atomic molecules \cite{7}. As the main application of quantum groups, q-deformed quantum mechanics \cite{8,9} was developed by generalizing the standard quantum mechanics which was based on the Heisenberg commutation relation (the Heisenberg algebra). Furthermore, quantum q-analogues of several fundamental notions and models in quantum mechanics have been mentioned such as phase space \cite{10}, uncertainty relation \cite{11,12}, density matrix \cite{13}, harmonic oscillator \cite{1,2,10}, hydrogen atom \cite{14}, creation and annihilation operators and coherent states \cite{1,2,15,16,17,18} so that they reduce to their standard counterparts as $q\rightarrow 1$. It can also be interpreted that some properties of generalized q-variables are notably different from the properties of the standard quantum mechanics because of imposing the deformation.\par

Recently, some efforts have been made to solve various problems of quantum mechanics by the Lie algebraic methods. These methods have been the subject of interest in many fields of physics and chemistry. For example these methods provide a way to obtain the wave functions of potentials in nuclear and polyatomic molecules \cite{19}. On the other hand, the deformed algebras are deformed versions of the usual Lie algebras which are obtained by introducing a deformation parameter q. The deformed algebras provide appropriate tools for describing systems which cannot be described by the ordinary Lie algebras. with this motivation,we studied the q-deformed Morse and harmonic potential in the nonrelativistic time-independent Schr\"odinger equation.\\
The present paper is organized as follows. In sect. 2, a  kind of deformation of quantum mechanics is introduced. Then as first study in such deformation, q-deformed Morse system has been investigated in Sect. 3. Furthermore, q-deformed harmonic oscillator is studied in Sec. 4 and finally, concluding remarks are given in sect. 5.

\section{The Classical Oscillator Algebra}

The classical oscillator algebra is defined by the canonical commutation relations \cite{18,19}:
\begin{equation}\label{1}
\left[ {{\rm{a,}}\,{{\rm{a}}^\dag }} \right]{\rm{  = 1,}}\,\,\,\,\,\,\,\,\,\,\,\left[ {N,a} \right] =  - a,\,\,\,\,\,\,\,\,\,\,\left[ {N,{a^\dag }} \right] = {a^\dag },
\end{equation}
where $N$ is called a number operator and it is assumed to be hermitian. The first deformation was accomplished by Arik and Coon \cite{22} as follows:

\begin{equation}\label{2}
a{a^\dag } - q{a^\dag }a = 1,\,\,\,\,\,\,\,\,\,\left[ {N,\,\,{a^\dag }} \right] = {a^\dag },\,\,\,\,\,\,\,\,\,\,\,\left[ {N,\,a} \right] =  - a
\end{equation}
where the relation between the number operator and step operators becomes
\begin{equation}\label{3}
{a^\dag }a = {\left[ N \right]_q},
\end{equation}
where a q-number is defined as
\begin{equation}\label{4}
{\left[ X \right]_q} = \frac{{1 - {q^X}}}{{1 - q}}
\end{equation}

The Jackson derivative is defined as \cite{23}
\begin{equation}\label{5}
\partial _x^qf\left( x \right) = \frac{{f\left( x \right) - f\left( {qx} \right)}}{{\left( {1 - q} \right)x}}
\end{equation}
which reduces to the ordinary derivative when $ q \rightarrow 1 $. If we introduce the coordinate realization of the deformed momentum ${\partial _q} = \frac{\hbar }{i}{\rm{ }}\partial _x^q$, we can obtain the q-deformed Schr\"odinger equation of the form

\begin{equation}\label{6}
\left[ { - \frac{{{\hbar ^2}}}{{2m}}\left( {\partial _x^q} \right)^2 + U\left( x \right)} \right]\psi \left( x \right) = E\psi \left( x \right)
\end{equation}

But, this equation is not easily solved for some potentials which has an analytic solutions in the limit $ q \rightarrow 1 $. Recently, another type of q-deformed theory appeared in statistical physics, which was first proposed by Tsallis \cite{24,25}. He used the q-deformed logarithm instead of an ordinary logarithm in defining the entropy called a Tsallis entropy. 


\section{ The q-deformed Morse Potential}
In this section, we demonstrate the viability of the canonical algebra as discussed in Section 2 for obtaining energy eigenvalue and eigenfunctions for q-deformed Morse potential.\\
The time-independent Schr\"odinger equation is given by
\begin{equation}\label{10}
\left( { - \frac{{{\hbar ^2}}}{{2m}}D_x^2 + V\left( x \right)} \right)U\left( x \right) = EU\left( x \right),
\end{equation}
where
\begin{equation}\label{11}
{D_x} = \left( {1 + q\left| x \right|} \right)\frac{\partial }{{\partial x}}.
\end{equation}
By using changing variable $X = \frac{1}{q}\ln \left( {1 + q\left| x \right|} \right)$, Eq. \eqref{10} reduces to:
\begin{equation}\label{12}
\left( { - \frac{{{\hbar ^2}}}{{2m}}\frac{{{d^2}}}{{d{X^2}}} + V\left( X \right)} \right)U\left( X \right) = EU\left( X \right).
\end{equation}
Now considering the q-deformed Morse potential \cite{26,27} of the form
\begin{equation}\label{13}
V\left( X \right) = {D_e}\left( {{e^{ - 2\beta \left( {X - {X_0}} \right)}} - 2{e^{ - \beta \left( {X - {X_0}} \right)}}} \right)
\end{equation}
and by putting $s = {e^{ - \beta \left( {X - {X_0}} \right)}}$, Eq. (\ref{12}) becomes
\begin{equation}\label{14}
\frac{{{d^2}U\left( s \right)}}{{d{s^2}}} + \frac{1}{s}\frac{{dU\left( s \right)}}{{ds}} + \frac{1}{{{s^2}}}\left( { - \frac{{2m{D_e}}}{{{\beta ^2}{\hbar ^2}}}{s^2} + \frac{{4m{D_e}}}{{{\beta ^2}{\hbar ^2}}}s + \frac{{2mE}}{{{\beta ^2}{\hbar ^2}}}} \right)U\left( s \right) = 0
\end{equation}
 Now solving Eq. (\ref{14}) eigenfunction can be written as

\begin{equation}\label{15}
U \left( X \right) = {e^{ - \beta \left( {X - {X_0}} \right)\sqrt {\frac{{ - 2mE}}{{{\beta ^2}{\hbar ^2}}}} }}{e^{ - \sqrt {\frac{{2m{D_e}}}{{{\beta ^2}{\hbar ^2}}}} \,{e^{ - \beta \left( {X - {X_0}} \right)}}}}L_n^{2\sqrt {\frac{{ - 2mE}}{{{\beta ^2}{\hbar ^2}}}} }\left( {2\sqrt {\frac{{2m{D_e}}}{{{\beta ^2}{\hbar ^2}}}} {e^{ - \beta \left( {X - {X_0}} \right)}}} \right)
\end{equation}
and the energy eigenvalue is
\begin{equation}\label{16}
{E_n} =  - {D_e} - \frac{{{\beta ^2}{\hbar ^2}}}{{8m}}{\left( {2n + 1} \right)^2} + \frac{{{\beta ^2}{\hbar ^2}}}{{2m}}\left( {2n + 1} \right)\sqrt {\frac{{2m{D_e}}}{{{\beta ^2}{\hbar ^2}}}}
\end{equation}

\section{ The q-deformed harmonic oscillator}

The q-deformed harmonic oscillator \cite {16,28}is given by
\begin{equation}\label{16a}
V\left( X \right) = \frac{1}{2}m{\omega ^2}{X^2}
\end{equation}
where $m$ and $\omega$ are the mass and frequency of oscillator.\\
Now substituting Eq.(\ref{16a}) into Eq. (\ref{12}) and by using changing variable $s = \frac{{m\omega }}{\hbar }{X^2}$ we obtain
\begin{equation}\label{17}
s\frac{{{d^2}U\left( s \right)}}{{d{s^2}}} + \frac{1}{2}\frac{{dU\left( s \right)}}{{ds}} + \left( {\frac{E}{{2\hbar \omega }} - \frac{s}{4}} \right)U\left( s \right) = 0
\end{equation}
For further convenience, we apply the gauge transformation
$U\left( s \right) = {e^{ - {s \mathord{\left/
				{\vphantom {s 2}} \right.
				\kern-\nulldelimiterspace} 2}}}\chi \left( s \right)$ which leads to
\begin{equation}\label{18}
s\frac{{{d^2}\chi \left( s \right)}}{{d{s^2}}} + \left( {\frac{1}{2} - s} \right)\frac{{d\chi \left( s \right)}}{{ds}} + \left( {\frac{E}{{2\hbar \omega }} - \frac{1}{4}} \right)\chi \left( s \right) = 0
\end{equation}

The Eq. (\ref{18}) is identified as the Kummer differential equation. In view of the above equations, the even and odd eigenfunctions may be, respectively, expressed as \cite{29}
\begin{subequations}\label{19}
\begin{align}
{U_{even}} = {N_n}{e^{ - \frac{{m\omega }}{{2\hbar }}{X^2}}}{H_n}\left( {X\sqrt {\frac{{m\omega }}{\hbar }} } \right)\\
{U_{odd}} = {N_n}{e^{ - \frac{{m\omega }}{{2\hbar }}{X^2}}}{H_n}\left( {X\sqrt {\frac{{m\omega }}{\hbar }} } \right)
\end{align}
\end{subequations}
where ${N_n}$ is the normalization constant. However, the even and odd eigenfunctions may be combined and the stationary
states of the relativistic oscillator are
\begin{equation}\label{20}
{U_n}\left( x \right) = {N_n}{e^{ - \frac{{m\omega }}{{2\hbar }}{{\left( {\frac{1}{q}\ln \left( {1 + q\left| x \right|} \right)} \right)}^2}}}{H_n}\left( {\frac{1}{q}\ln \left( {1 + q\left| x \right|} \right)\sqrt {\frac{{m\omega }}{\hbar }} } \right)
\end{equation}
The energy eigenvalues of spin-zero particles bound in this oscillator potential may be found using (\ref{17}). Therefore, the energy for even and odd states can be written as

\begin{equation}\label{21}
{E_n} = \left( {n + \frac{1}{2}} \right)\hbar \omega
\end{equation}
where $n = 0, 1, 2, 3,..... $ are the integers.  Note that Eq. (\ref{21}) is in agreement with the energy of the harmonic oscillator.

\section{Conclusion}
In this paper, we have introduced the q-deformed morse and harmonic oscillator potential functions in the light of canonical commutation algebra. we have computed the energy eigenvalues and corresponding eigenfunctions for these potentials in one-dimensional nonrelativistic Schr\"odinger equation. The exact solution for the eigenfunction is obtained in terms of Laguerre polynomial for Morse potential. However, in case of harmonic oscillator, even and odd eigenfunctions are obtained in terms of Hermite polynomials. It is also noted that the function behavior depends on the variation of q-deformed parameter as well as the strength of the potential parameter. Moreover, it is worth to mention that in the limit case, results in ordinary quantum mechanics can be recovered.

\end{document}